\documentstyle[prl,aps,epsf,multicol]{revtex}
\begin{document}
\draft
\title{Statistics of Atmospheric Correlations}
\author{M. S. Santhanam $^{*}$ and Prabir K Patra$^{\dagger}$}
\address{IBM-Research, India Research Laboratory, Block-1, Indian
Institute of Technology,\\
New Delhi 110 016, India.}
\date{\today}
\maketitle
\begin{abstract}

For a large class of quantum systems the statistical properties
of their spectrum show remarkable agreement with random matrix
predictions. Recent advances show that the scope of random matrix
theory is much wider. In this work, we show that the random matrix
approach can be beneficially applied to a completely different classical
domain, namely, to the empirical correlation matrices obtained from
the analysis of the basic atmospheric parameters that characterise
the state of atmosphere. We show that the spectrum of atmospheric
correlation matrices satisfy the random matrix prescription.
In particular, the eigenmodes of the atmospheric empirical correlation
matrices that have physical significance are marked by deviations from
the eigenvector distribution.

\end{abstract}
\pacs{PACS number(s): 05.45.Tp, 92.70.Gt, 05.40.-a, 02.50.Sk \hskip 2.2cm
To appear in Phys. Rev. E}
%\narrowtext

\begin{multicols}{2}
\section{Introduction}
The study of random matrix ensembles have brought in a great deal of
insight in several fields of physics ranging from nuclear, atomic
and molecular physics,  quantum chaos and
mesoscopic systems \cite{wei}.
The interest in random matrices arose from
the need to understand the spectral properties
of the many-body quantum systems with complex interactions.
With general assumptions about the symmetry properties of the system
dictated by quantum physics,
random matrix theory (RMT) provides remarkably successful predictions
for the statistical properties of the spectrum,
which have been numerically and experimentally verified in the last
few decades \cite{kud}.
In recent times, it has been realised
that the fluctuation properties of low-dimensional systems, 
{\it e.g.} chaotic quantum systems, are universal and can be modelled
by an appropriate ensemble of random matrices \cite{boh}.
From its origins in quantum physics of high dimensional systems,
the scope of RMT is further widening with the new approaches
based on supersymmetry methods \cite{efe} and applications in 
seemingly disparate fields like quantum chromodynamics \cite{ver},
two-dimensional quantum gravity \cite{abd}, conformal field theory \cite{wei}
and even financial
markets \cite{lal}. Thus, random matrix techniques have potential applications
and utility in disciplines far outside of quantum physics.
In this, we show that the empirical correlation matrices
that arise in atmospheric sciences can also be modelled
as a random matrix chosen from an appropriate ensemble.

The correlation studies are elegantly carried out in the matrix framework.
The empirical correlation matrices arise in a multivariate setting
in various disciplines;
for instance, in the analysis of space-time data in 
general problems of image processing and pattern recognition, in particular
for image compression and denoising \cite{img};
the weather and climate data are frequently subjected to principal
component analysis to identify the independent modes
of atmospheric variability \cite{pre}; in the study of financial assets
and portfolios
through the Markowitz's theory of optimal portfolios \cite{mark}.
Most often, the analysis performed on the correlation matrices is
aimed at separating the signal from `noise', {\it i.e.} to cull the physically
meaningful modes of the correlation matrix from the underlying noise.
Several methods based on Monte-Carlo simulations have been
used for this purpose \cite{pre1}. The general premise of 
such methods
is to simulate 'noise' by constructing an ensemble of matrices with
random entries drawn from specified distributions and the statistical
properties of its eigenvalues, like
the level density etc., are compared with that of the correlation matrices. 
Even as the Monte-Carlo techniques become computationally expensive
beyond a point, asymptotic formulations take over. The deviations from
'pure noise' assumptions are interpreted as signals or symptom of
physical significance.
In the context of the atmospheric sciences, empirical
correlation matrices are widely used, for example, to study the large scale
patterns of atmospheric variability. If the random matrix
techniques are valid for a correlation matrix, it might be
useful as a tool to separate the signal from the noise, with lesser
computational expense than with methods based on Monte-Carlo techniques.
We show that RMT prediction for eigenvector distribution has
potential application in this direction for atmospheric
correlation matrices.

\section{Correlations and Teleconnections}

The state of the atmosphere is governed by the classical laws of
fluid motion and exhibits a great deal of correlations in
various spatial and temporal scales. These correlations
are crucial to understand the short and long term trends in climate.
Generally, atmospheric correlations can be recognised from the study of
empirical correlation matrices constructed using the atmospheric data.

\end{multicols}

\newpage
\widetext

\begin{figure}
\widetext
\epsfxsize 6in
\centerline{
\epsfbox{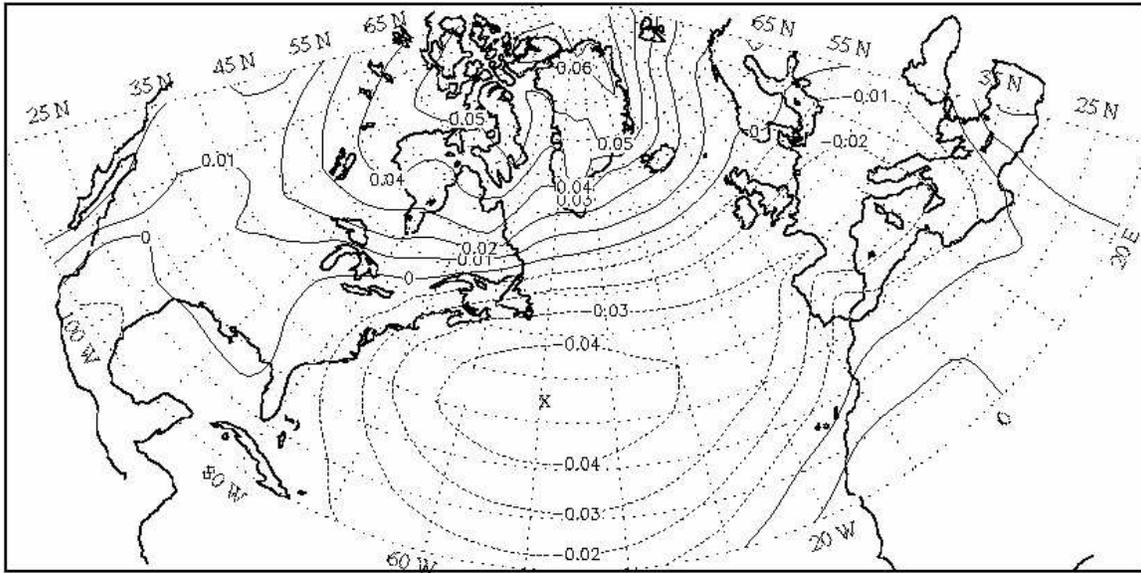}}
\widetext{
\caption{The NAO pattern as captured by the EOF analysis of sea-level
pressure, with the geographical map of the domain of analysis in the
background. The contours are drawn after averaging over the first two
dominant EOFs. Note the north-south dipole shown as closed contours,
in mid-Atlantic (dotted contour) and over Greenland (solid contours).}
\label{naoeof}
}
\end{figure}

\vskip -5in
%\widetext

\begin{figure}
%\widetext
%\epsfysize 6in
\centerline{
\epsfbox{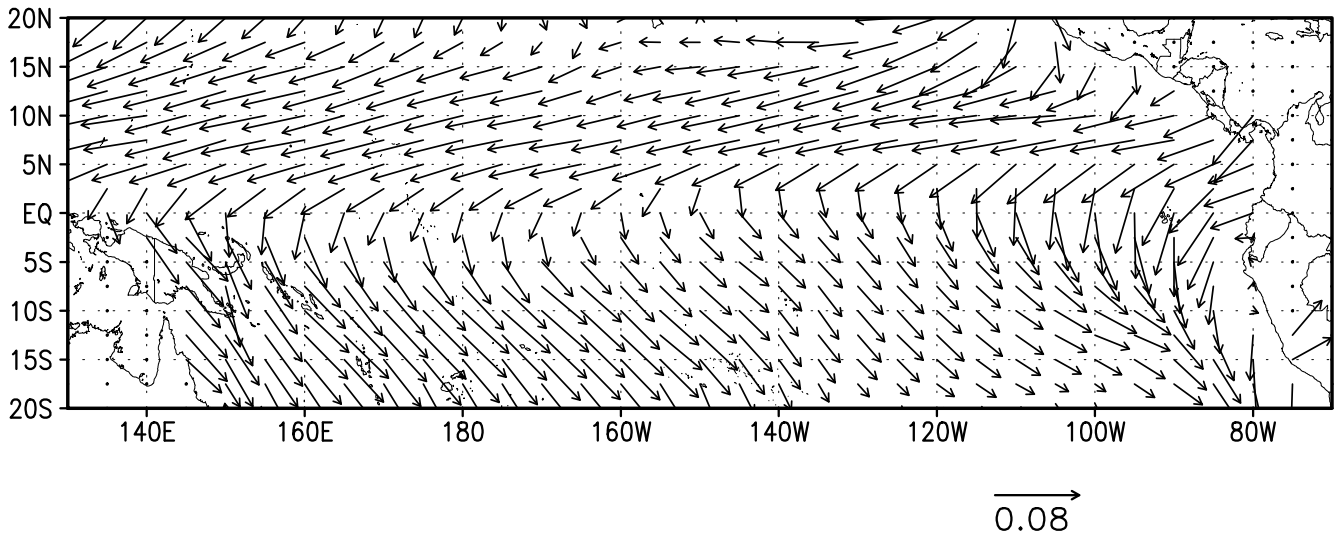}}
\vskip -0.5cm
\widetext{
\caption{Dominant EOF from analysis of wind-stress vectors with the equatorial
Pacific Ocean domain in the background. This eigenmode predominantly represents
the annual fluctuations in trade winds and accounts for 38\% of the variability.
This eigenvector has been rotated by $45^o$ to obtain physically meaningful
pattern.}
\label{ceof}}
\end{figure}

\newpage

\begin{multicols}{2}

Most significant correlations
are documented as teleconnection patterns, {\it i.e.}, the simultaneous
correlations in the fluctuations of the large scale atmospheric
parameters at widely separated points on the earth.
They could be thought of as the dominant modes of atmospheric variability.
Wallace and Gutzler have surveyed the entire northern hemisphere
teleconnections and show that the dominant eigenmodes of the
correlation matrices, in most cases,
reflect these teleconnection patterns \cite{wal}.
For instance, the North Atlantic Oscillation (NAO) \cite{nao} refers to the
exchange of the atmospheric mass between Greenland/Iceland region
and the regions of North Atlantic ocean between $35^o\mbox{N}$ and
$40^o\mbox{N}$ and is characterised by a north-south dipole pattern as
shown in Fig. \protect\ref{naoeof}.
It is known that the NAO is associated
with anomalous weather patterns in the eastern US and
northern Europe including Scandinavia \cite{jam}.
Such dominant modes need not always have
to be a teleconnection. For example, the pattern in Fig \protect\ref{ceof} can
be identified with the annual trade wind fluctuations in
the equatorial Pacific region; obtained as a dominant
eigenmode from the analysis of the pseudo wind stress
vectors. In subsequent sections, we will perform statistical analysis on the
spectra of atmospheric correlation matrices, whose dominant
modes display correlation patterns discussed above.
Atmospheric correlations are interesting to study from a RMT perspective because
they arise naturally from known physical interactions and offers
instances to verify two (orthogonal and unitary) of the three Gaussian
ensembles of RMT.

\subsection{Empirical Orthogonal Functions}
The Empirical Orthogonal Function (EOF) method, also called the
Principal Component Analysis,
is a multivariate statistical technique widely used in the analysis
of geophysical data \cite{pre}. It is similar to the singular value
decomposition employed in linear algebra and it provides
information about the independent modes of variabilities exhibited 
by the system. 

In general, any atmospheric
parameter $z(x,t)$, (like wind velocity, geopotential height, 
temperature etc.),
varies with space$(x)$ and time$(t)$ and is assumed
to follow an average trend on which the variations (or anomalies, as
referred to in atmospheric sciences) are superimposed, {\it i.e.},
$ z(x,t) = z_{avg}(x) + z'(x,t) $.
The wind vectors can be represented as a complex number,
$s e^{i \theta}$  where $s$ is the wind speed and $\theta$ the
direction. Thus, in general, $ z(x,t)$ could be a complex number. The
mathematical treatment of complex correlations and
EOFs is given in ref \cite{har}.
In further analysis, the standardised anomaly $ z'(x,t) $ will be used
which will have zero mean
($\overline{z'}(x) = 0 $) and is rescaled such that
its variance $ < z'(x)^2 > $ is unity.
If the observations were taken $n$ times at each of the $p$ spatial locations
and the corresponding anomalies $z'(x,t)$ assembled in the data 
matrix $\mbox{\bf Z}$ of order $p$ by $n$,
then the spatial correlation matrix of the anomalies is given by,
\begin{equation}
\mbox{\bf S} = \frac{1}{n} \; \mbox{\bf Z} \; \mbox{\bf Z}^{\dagger}
\label{cmat}
\end{equation}
Note that the elements of the hermitian matrix $\mbox{\bf S}$, 
of order $p$, are just the Pearson correlation between various
spatial points.
The eigenfunctions of $\mbox{\bf S}$ are called the empirical 
orthogonal functions since they form a complete set of orthogonal basis
to represent the data matrix $\mbox{\bf Z}$.
In the geophysical setting, the EOFs can be plotted as contour maps
by associating each component with its corresponding spatial location as
shown in Fig \protect\ref{naoeof}. If the eigenvalue corresponding
to the $m$th eigenmode is $\lambda_m$, then the
percentage variance associated with the mode is
given by, $ v_m = (\lambda_m/\sum^p_{i=1} \lambda_i) 100.0.$
Then, the dominant mode
would correspond to the EOF with the largest eigenvalue. 
In the last few decades, several variants of this basic EOF technique
have been used to suit varied requirements \cite{pre}.
We will show that the spectrum of $\mbox{\bf S}$ displays random
matrix type spectral statistics.

\section{Eigenvalue statistics}
\subsection{Data and analysis}
Computing reliable correlation matrices depend on the availability
of sufficiently long time series of data. Generally, the requirement
is to have $ n >> p$, as otherwise the computed covariances
could be noisy and correlations could be regarded as random. 
Reliable records of monthly averages for weather
and climate parameters of interest exist for the last 50 years.
In our study, we use both
the daily as well as the monthly averaged data available
from NCEP reanalysis archives \cite{data}.
Further in this direction, we study three cases;
(i) monthly mean sea level pressure (SLP) for the Atlantic domain
$(0-90^o\mbox{N}, 120^o\mbox{W} - 30^o\mbox{E})$ from 1948 to 1999.
(ii) monthly mean global sea surface temperatures (SST) \cite{rey} and
(iii) surface level pseudo wind-stress vectors in the equatorial Pacific ocean
$(20^o\mbox{S}-20^o\mbox{N}, 130^o\mbox{E}-70^o\mbox{W})$.
The first case identifies many northern hemisphere teleconnections
and its climatic effects and EOF aspects
are documented \cite{wal}. Wind-stress is an
important quantity in studies on coupled ocean-atmosphere models
that simulate the air-sea interaction and the feedback mechanism.
The pseudo wind-stress is defined as $ W = \sqrt{(u^2+v^2)} ( u + i v ) $,
where $u$ and $v$ are the zonal and meridional wind components, and
this leads to complex correlation matrix. Its EOFs exhibit
signatures of the mean annual signal and ElNino oscillations \cite{leg}.
Note that the eigenmodes of complex correlation matrix
are determined
only up to a complex factor of unit modulus. This allows the freedom
to choose a phase angle to rotate the eigenvectors.

The atmospheric data is on an uniform spatial grid
of $2.5^o$  along both the latitude and longitude.
To ensure that $n > p$, in the calculations with monthly mean data, 
the spatial resolution was reduced to $5^o$.
Thus, for the case(i) of monthly mean SLP correlations,
p=434 and n=624. In the case (iii) of monthly mean wind stress analysis
over equatorial Pacific ocean,
the land points were removed from the calculations using land-sea mask
and it results in p=494 and n=624.
Since a longer time-series of monthly mean data was not available,
another experiment was performed with daily averaged time-series from
1990 to 1999 with much improved ratio for $r = n/p$ in the
range 2.5-3.5. The required means and anomalies were computed
from which matrices of orders ranging from
500 to 1200 is constructed and diagonalised using standard
LAPACK routines \cite{lap}.

\begin{figure}
\epsfxsize 3.0in
\centerline{
\epsfbox{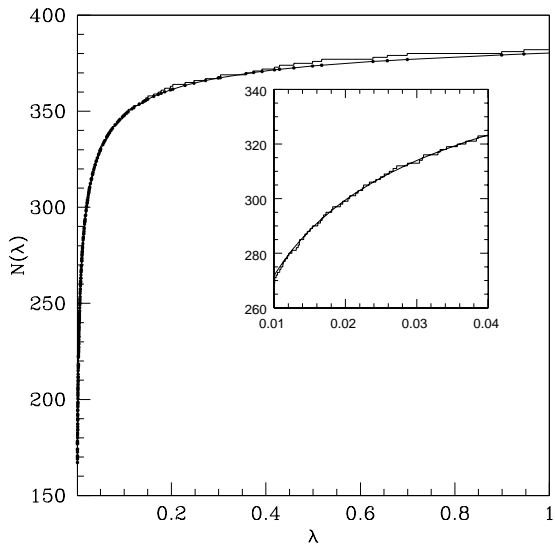}}
\narrowtext{
\caption{The integrated level density, in the form of a staircase,
for the eigenvalues of the monthly mean SLP correlation matrix.
The solid line is the empirical curve that fits the level density
with dark circles denoting the location of eigenvalues.
The inset is magnified view of a part of the curve.}
\label{dos}
}
\end{figure}

First we look at the structure of eigenvalue density.
The integrated level density 
$ N(\lambda) = \sum \Theta(\lambda - \lambda_i)$,
can be written as, $ N(\lambda) \approx N_{avg}(\lambda) + N_{fl}(\lambda) $,
a sum of an average part and the fluctuating part.
The eigenvalues $\lambda_i$
are unfolded by fitting an empirical function to the average part of
the integrated level density such that the unfolded eigenvalues
$\epsilon_i = N_{avg}(\lambda_i)$ have
mean spacing unity \cite{boh1}. All the analysis reported further were
performed on $\epsilon_i$.
As Fig. \protect\ref{dos} shows, for empirical correlation matrices, the spectrum
is dense at the lower end. This is typical of the spectrum of
correlation matrices formed from the data matrix $\mbox{\bf Z}$ through 
eq (\protect\ref{cmat}) \cite{mitra}.  In contrast to this,
for a generic quantum system, the level density increases
with energy and is dense at the higher end of the spectrum.

\subsection{Level spacing distribution}
One of the celebrated results of the random matrix theory is the
nearest-neighbour eigenvalue spacing distribution; {\it i.e.} the distribution
of $s_i = \epsilon_{i+1}-\epsilon_i$. It gives the probability for finding
the neighbouring levels with a given spacing $s$.
In the context of this work,
the Gaussian Orthogonal Ensemble (GOE) is appropriate for the
mean sea-level pressure correlations and Gaussian Unitary Ensemble
(GUE) for pseudo wind-stress vectors. The spectra of these classes
exhibit universal fluctuation properties and the spacing distributions
are given by \cite{meh},
\begin{eqnarray}
P_{\mbox{\tiny GOE}}(s) & = & \frac{\pi}{2} \; s \exp(- \frac{\pi}{4} s^2 ) \label{evs1} \\
P_{\mbox{\tiny GUE}}(s) & = & \frac{32}{\pi^2} \; s^2 \exp(- \frac{4}{\pi} s^2 ) \label{evs2}
\end{eqnarray}
The analytical forms above indicate level-repulsion, a tendency against
clustering, as evident from low probability for small spacings. The
level repulsion is linear for GOE and quadratic for GUE.

\begin{figure}
\epsfxsize 3.0in
\centerline{
\epsfbox{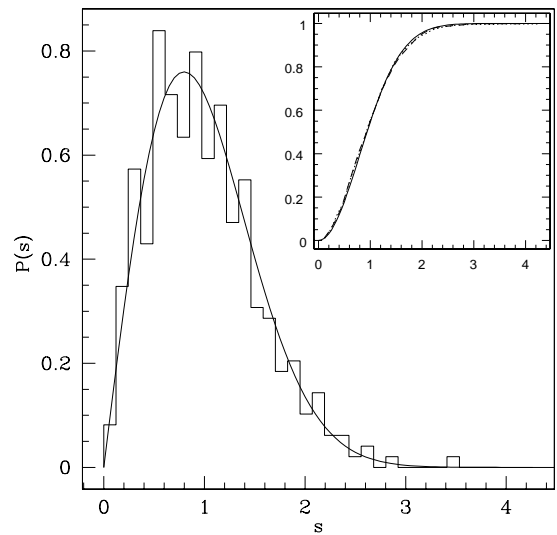}}
\narrowtext{
\caption{Eigenvalue spacing distribution for
the monthly mean SLP correlation matrix. The solid curve is the GOE prediction.
The inset shows the cumulative distribution for the monthly and daily averaged
correlation matrix.}
\label{evalgoe}
}
\end{figure}

\begin{figure}
\epsfxsize 3.0in
\centerline{
\epsfbox{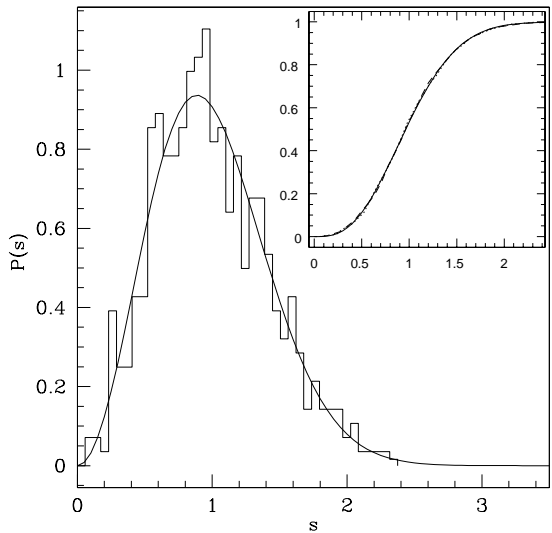}}
\narrowtext{
\caption{Eigenvalue spacing distribution for
the monthly mean wind-stress correlation matrix. The solid curve is the GUE prediction.
The inset shows the cumulative distribution for the monthly and daily averaged
correlation matrix.}
\label{evalgue}
}
\end{figure}

In Fig. \protect\ref{evalgoe}, we show the spacing distribution for the 
eigenvalues of the
correlation matrix of the monthly mean SLP. The inset in this
figure shows the cumulative spacing distribution for the spectra obtained
from the analysis of monthly and daily averaged SLP data. We observe a
general agreement with the RMT predictions. In Fig. \protect\ref{evalgue}, the spectra from
the monthly mean wind-stress correlation data is shown. If the spacings, $s$,
were uncorrelated then we would expect a Poisson distribution, $P(s) = 
\exp(-s)$ \cite{boh1}. In all the cases we studied, the empirical histograms 
do not follow the Poisson curves at all. As would be
expected, a better agreement between the theoretical curves and the empirical
distributions is observed in the analysis of daily averaged data,
in both the cases of SLP and pseudo wind-stress correlations,
since they provide about 1000 eigenvalues for the statistics.
For instance, a Kolmogrov-Smirnov test at 65\% confidence level could not 
reject the hypothesis that GOE is the correct distribution for the eigenvalues
of monthly mean SLP correlation matrix, whereas a similar test for the daily 
averaged SLP data could not reject the hypothesis at 99\% confidence level.
The monthly mean SST correlation matrix analysis (not shown here) also
supports RMT spacing distribution.
The eigenvalue spacing distribution for the
equatorial Pacific pseudo wind-stress vector correlation matrix also
indicates a good agreement with the GUE prediction given by 
Eq (\protect\ref{evs2}) (see Fig \protect\ref{evalgue}). 

\subsection{Long-range Correlations}
Beyond the nearest-neighbour spacing distribution, we study the long-range
correlations. We compute the following spectral fluctuation measures \cite{boh1} which
are based on the two-point correlation function.
(a) The spectral rigidity, the so-called $\Delta_3$ statistic,
measures
the least-square deviation of the spectral staircase function $N(\epsilon)$
from the straight line of best fit for a finite interval $L$ of the spectrum,
\begin{equation}
\Delta_3(L,L') = \frac{1}{L} \stackrel{\mbox{min}}{a,b} \int^{L'+L}_{L'}
[ N(\epsilon) - a \epsilon - b]^2 \;\; d\epsilon
\end{equation}
where $a$ and $b$ are obtained from a least-squares fit.
Average over several choices of $L'$ gives the spectral rigidity $\Delta_3(L)$.
(b) the number variance $\Sigma^2$ is also a function of two-point
correlation function. Let $n(L, L')$ be the number of eigenvalues
in the spectral interval $L$. Then, for a choice of $L'$, $\Sigma^2$ is given by,
\begin{equation}
\Sigma^2(L,L') =  n(L,L')^2 - L^2
\end{equation}
Averaging $ n(L,L')^2 $ over $L'$ gives the number variance $\Sigma^2(L)$.
The asymptotic results, for large $L$, from random matrix considerations, is given by
\cite{meh},
\begin{eqnarray}
\Delta_3(L) & = & \frac{1}{\nu \pi^2} \log(L) + g_{\nu} \\
\Sigma^2(L) & = & \frac{2}{\nu \pi^2} \log(L) + h_{\nu}
\end{eqnarray}
where $\nu=1,2$ corresponds to GOE and GUE respectively; $g$ and $h$
are also dependent on the ensemble.

\begin{figure}
\epsfxsize 3.0in
\centerline{
\epsfbox{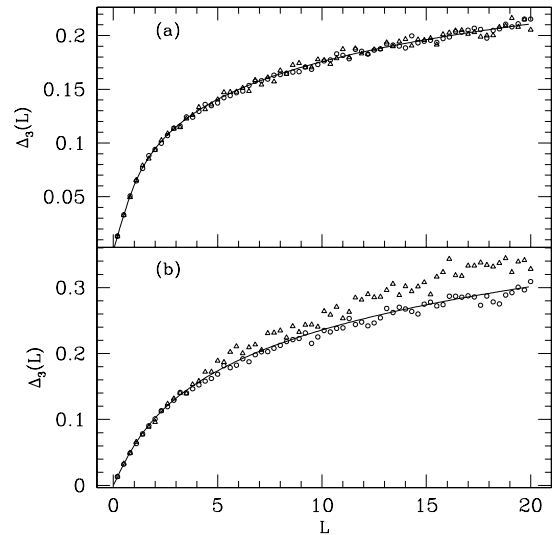}}
\narrowtext{
\caption{$\Delta_3(L)$ for spectra from the correlation matrix of
(a) wind stress and (b) SLP. The solid curve in (a) is GUE prediction
and in (b) the GOE prediction.
The circles are for the correlation matrix obtained from daily averaged
data and triangles represent the matrix obtained from the monthly mean data.}
\label{del3}
}
\end{figure}

\begin{figure}
\epsfxsize 3.0in
\centerline{
\epsfbox{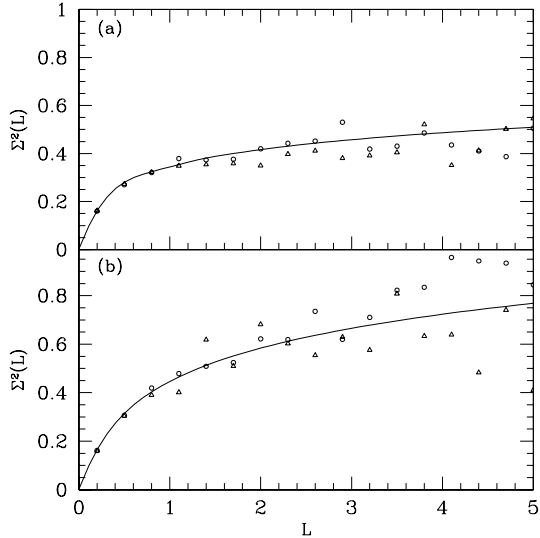}}
\narrowtext{
\caption{$\Sigma^2(L)$ for spectra from the correlation matrix of
(a) wind stress and (b) SLP.
The circles are for the correlation matrix obtained from daily averaged
data and triangles represent the matrix obtained from the monthly mean data.}
\label{sig2}
}
\end{figure}

Fig \protect\ref{del3} shows the $\Delta_3(L)$ statistic for the SLP and wind-stress
correlation matrix spectrum, computed using the method
given by Bohigas and Giannoni \cite{boh1}.
Generally, a good agreement is observed with the RMT predictions.
In all the cases, for small $L$ the agreement is good and small deviations begin
to be seen for larger values indicating a possible breakdown of universality.
In general, this should indicate system specific features that cannot be
modelled by assumptions based on randomness.
Once again, we notice that the correlation matrix spectra obtained from daily
data show better agreement with RMT predictions, primarily due to larger
orders of correlation matrix involved and hence more eigenvalues for the analysis.
Fig \protect\ref{sig2} shows the number variance $\Sigma^2(L)$ for all the cases.
We observe a
fairly good agreement with RMT predictions. The results for SLP and SST
correlations are in broad agreement with the similar analysis performed
on the financial correlation matrices \cite{lal}, both of
which are modelled by the orthogonal ensemble of RMT. This, in itself,
demonstrates the breadth of applications of RMT.

\section{Statistics of EOF components}
With the eigenvalue statistics, it is not straightforward to obtain 
detailed system
specific information, unless there are significant deviations from
random matrix predictions. 
The distribution of eigenvector components, on the other hand, reveals
fine-grained information, at the level of every eigenvector.
In this section, we show that almost all the EOFs follow the
RMT distribution.
However, a few EOFs that have physical
significance, like the ones shown in Figs \protect\ref{naoeof} and \protect\ref{ceof}, 
deviate strongly from RMT.
Broadly, the
variability captured by an EOF is seen to be reflected in its
deviation from RMT predictions.

Let $ a^m_j $ be the $j$th component of the $m$th eigenvector. Assuming that
these components are Gaussian random variables with the norm being their only
characteristic, it can be shown that the distribution of $ r = | a^m_j|^2 $,
in the limit when the matrix dimension is large,
is given by the special cases of the $\chi^2$ distribution \cite{hake},
\begin{equation}
P_\nu (r) = \left( \frac{ \nu}{2 <r>} \right)^{\nu/2} \;
\frac{r^{\nu/2 -1}}{\Gamma(\frac{\nu}{2})} \;
\exp\left(\frac{-r \nu}{2 <r>} \right)
\end{equation}
The case $\nu=1$ can be identified with GOE and gives the well-known
Porter-Thomas (PT) distribution. The distribution of complex eigenvectors
correspond to GUE class with $\nu=2$.  The general understanding is that if
the eigenvectors are sufficiently irregular in some sense, then its components
are $\chi^2$ distributed and deviations occur if they show some symptoms
of regularity.

\begin{figure}[t]
\epsfxsize 3.0in
\centerline{
\epsfbox{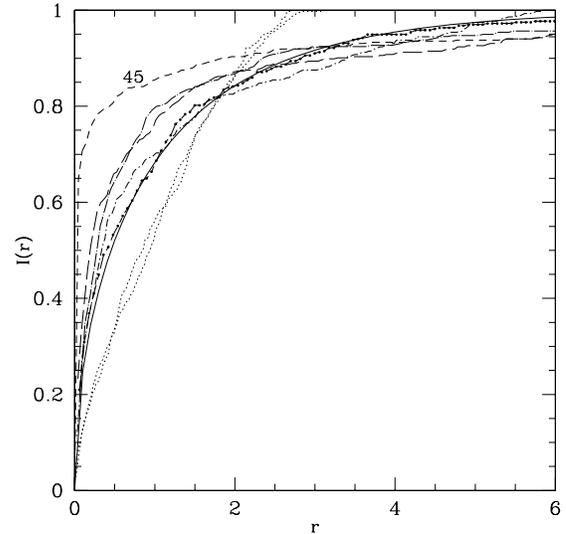}}
\narrowtext{
\caption{Cumulative distribution of EOF components for the SLP correlation matrix.
The solid curve is the Porter-Thomas distribution. The two curves with dotted lines
correspond to the first two dominant EOFs, whose spatial map is shown in
Fig. \protect\ref{naoeof}. The long-dashed curves are the next few dominant
EOFs. The curve with dark-circles is a sample from the bulk of lesser-dominant EOFs
that mostly follow Porter-Thomas distribution. The curve with small dashes (marked as 45)
is the 45th EOF that surprisingly deviates from the PT curve (see text).}
\label{evecgoe}
}
\end{figure}

In further analysis, we will
use the modulus square of the EOF components, {\it i.e.} $r = |a^m_j|^2$,
normalised to unit mean.
For the monthly mean SLP correlation matrix, Fig \protect\ref{evecgoe} shows 
the cumulative distribution of EOF components.
Since EOFs form an optimal basis to represent the
data, most of the variability is carried by a small number of EOFs; in this
case about 91\% of the variability is captured by just
12 dominant EOFs. The rest 9\% is accounted for by the bulk of
the rest 422 EOFs. The central result of this section is that the
bulk of these EOFs, accounting for a small fraction
of the variability, follow the cumulative Porter-Thomas (PT) distribution
given by, $I(r) = \mbox{erf}(\sqrt{r/2})$, where $\mbox{erf}$ is the standard
error function. This strengthens
the conclusion that the empirical correlation matrices can be modelled
as a random matrix. As an example from a large number of
such EOFs, the distribution of 294th EOF is shown (denoted by
dark circles) in Fig \protect\ref{evecgoe}, and it practically
falls on the PT curve. We observed that the distribution of the
all such EOFs follow RMT and this is also confirmed
by a Kolmogrov-Smirnov test.

\begin{figure}
\epsfxsize 3.0in
\centerline{
\epsfbox{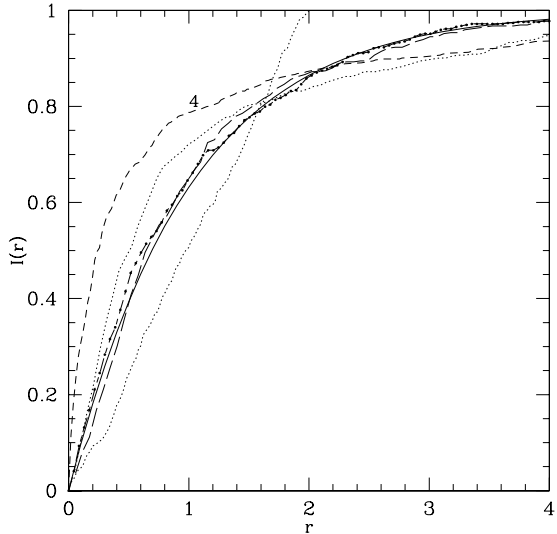}}
\narrowtext{
\caption{Cumulative distribution of EOF components for the pseudo wind-stress
correlations. The solid curve is the GUE prediction. The dotted curves are for
the first two dominant EOFs; the spatial map of dominant EOF shown in
Fig. \protect\ref{ceof}. The curve with dark-circles is a sample from the bulk
of lesser-dominant EOFs that follow GUE. The curve with small dashes
(marked as 4) is the 4th EOF deviates significantly from the GUE prediction.}
\label{evecgue}
}
\end{figure}

However, interesting cases arise from a small number of dominant EOFs which
deviate strongly from RMT predictions.
The first two dominant EOFs shown in Fig \protect\ref{evecgoe}
(as dotted lines),
representing about 30\% and 22\% of the entire variability, show significant
deviations from cumulative PT curve.
The spatial structure of both these eigenmodes, shown in Fig \protect\ref{naoeof},
jointly capture the essence of North-Atlantic pattern.
This scenario, of most dominant of the
EOFs deviating from the PT distribution and lesser
significant ones showing agreement with it, is repeated
in the analysis of SST (not shown here) and daily averaged SLP correlations
as well. At this point, we stress that these deviations are exceptions
that arise in about 1\% of the EOFs.

Fig \protect\ref{evecgue} shows the cumulative distribution for the EOFs obtained
from the analysis of the monthly mean wind-stress correlation.
Note that in this case, the appropriate prediction follows the unitary
ensemble since the EOF components are complex. The dominant
20 EOFs explain nearly 90\% of the variability in the wind stress data.
The rest of the large number (about 400) of EOFs show good
agreement with cumulative GUE curve for eigenvector distribution
given by $I(r) = 1 - \exp(-r)$. One such case, 370th EOF, is shown in 
Fig \protect\ref{evecgue} denoted by dark circles. In general, EOFs show good
agreement with RMT except for the few dominant EOFs.
The dominant EOF, whose spatial pattern is shown in Fig \protect\ref{ceof},
represents the mean annual Pacific trade-wind fluctuations
and explains 38\% of
the variability and shows pronounced deviation from the cumulative GUE curve.
Next few dominant EOFs also exhibit significant deviations.
Legler \cite{leg} has performed EOF analysis on the Pacific ocean
wind-stress vectors and
attributed physical significance to the top three dominant EOFs.
Thus, EOFs that have physical significance, cannot be modelled by
RMT ensembles.
An analogy with quantum eigenstates seems inevitable. 
Studies on the distribution of the eigenfunctions of low-disorder
tight-binding systems and chaotic quantum systems
show that a small fraction of the eigenstates, which display
quantum localisation,
deviate from random matrix predictions \cite{mul},
while most others show RMT-like behaviour.

There are two interesting observations in this study. Firstly, we notice 
that there are few EOFs, occurring at irregular intervals, which do not
carry much of a significance in terms of the variability but deviate
strongly from RMT predictions. It is not immediately clear if they carry
any significant information. Secondly, a surprising observation is that
the EOFs, corresponding to first few eigenvalues at the lower end of the
spectrum, most often regarded as least dominant and random, devoid of any
system specific information, show marked deviations from RMT (see also
ref \cite{lal}). One such example for each of GOE and GUE case is shown
in Figs \protect\ref{evecgoe} and \protect\ref{evecgue}.

\section{Discussion and conclusion}
This work shows that the random matrix predictions are of considerable
interest in the study of the correlation matrices that arise in atmospheric
sciences. Previous work on the correlations of stock market fluctuations
has come to similar conclusion \cite{lal}. This is despite the following
basic difference; RMT assumes that the quantum Hamiltonian matrix is part
of an ensemble of random matrices whose entries are independent random numbers
drawn from a Gaussian distribution. In the correlation matrix formalism,
the elements of data matrix are independent Gaussian distributed random
numbers. Then, the correlation matrix in eq. \protect\ref{cmat} follows
Wishart structure \cite{wish}, a form of generalised $\chi^2$-distribution.

In the application of EOFs in various disciplines
an important question is the truncation of EOFs while opting for a
low-dimensional representation for a given data matrix. The earlier
approaches to this problem were based on Monte-Carlo techniques or asymptotic
theories \cite{pre,pre1}. It would be interesting to evolve
a truncation
criteria, for using EOFs as empirical basis, from random matrix techniques
since the results here suggest that RMT could be potentially applied
to separate the random modes from the
physically significant modes of the correlation matrix.

Even as we have documented evidence for RMT like behaviour from the
atmospheric correlation matrices, there is also a need to look at the limits
of RMT description.
For instance, a correlation matrix which shows perfect
correlation will obviously not behave like RMT. Can correlation matrix spectra
display Poisson spacing distribution ? Such limits of RMT in the
context of correlation matrix is yet to be explored.

In summary, we have analysed atmospheric correlation matrices from
the perspective of random matrix theory. The central result of this 
work is that they
can be modelled as random matrices chosen from an appropriate RMT
ensemble. The eigenvalue statistics exhibits
short and long-range RMT-type behaviour.  Most of the eigenmodes also
follow the RMT type eigenvector distribution.
Few dominant eigenmodes
that have physical significance deviate from RMT predictions.
We have verified our conclusions with examples
of correlation matrices that belong to GOE and GUE universality classes of 
random matrix theory.

\acknowledgments
The atmospheric data used in this work is the NCEP Reanalysis data
provided by NOAA-CIRES Climate Diagnostics Center, Boulder, Colorado, USA,
from their Web site at http://www.cdc.noaa.gov. NCEP/NOAA-CIRES is
thankfully acknowledged for the same. We thank Dr. Abhinanda
Sarkar for clarifications on the nuances of statistics.
\vskip 5mm
\noindent $^{*}$ msanthan@in.ibm.com \\
$^{\dagger}$ Now at Atmospheric Composition Research Programme,
Frontier Research System for Global Change, Yokohama 236-0001, Japan.
prabir@jamstec.go.jp

\end{multicols}

\end{document}